\documentclass[12pt]{article}
\usepackage{amssymb,amsmath,epsfig}

\begin{document}

\title{\bf Warm Anisotropic Inflationary Universe Model}
\author{M. Sharif \thanks {msharif.math@pu.edu.pk} and Rabia Saleem
\thanks{rabiasaleem1988@yahoo.com}\\
Department of Mathematics, University of the Punjab,\\
Quaid-e-Azam Campus, Lahore-54590, Pakistan.}

\date{}
\maketitle

\begin{abstract}
This paper is devoted to study the warm inflation using vector
fields in the background of locally rotationally symmetric Bianchi
type I universe model. We formulate the field equations, slow-roll
and perturbation parameters (scalar and tensor power spectra as well
as their spectral indices) under slow-roll approximation. We
evaluate all these parameters in terms of directional Hubble
parameter during intermediate and logamediate inflationary regimes
by taking the dissipation factor as a function of scalar field as
well as a constant. In each case, we calculate the observational
parameter of interest, i.e., tensor-scalar ratio in terms of
inflation. The graphical behavior of these parameters shows that the
anisotropic model is also compatible with WMAP7 and Planck
observational data.
\end{abstract}

\section{Introduction}

In the context of cosmology, it has become an inevitable fact
through the combined efforts of type Ia Supernova, the large scale
structure (LSS), cosmic microwave background (CMB) and WMAP that
universe is undergoing a stage of accelerating expansion \cite{1}.
This cosmic behavior might be due to the repulsive nature of missing
energy (due to the large negative pressure) called dark energy (DE)
which occupies 70 percent of the universe. It is described by a tiny
time-independent cosmological constant $(\Lambda)$ obeying
$\omega=-1$ ($\omega$ is the equation of state (EoS) parameter). Due
to the fine-tuning and cosmic coincidence issues \cite{4} of this
constant, the search for variety of DE models is ongoing. The
dynamical nature of DE is divided into two categories: scalar field
models (quintessence, phantom with negative kinetic term, k-essence,
quintom (unification of quintessence and phantom) etc.) \cite{5} and
interacting DE models (Chaplygin gas having novel EoS, holographic
DE, Ricci DE etc.) \cite{6}.

The standard cosmology (big-bang model) successfully explains the
observations of CMB radiations but still there are some unresolved
issues. The early universe is facing some long standing problems
including horizon problem (why is the universe isotropic and
homogeneous on large scale structure?), flatness (why is
$\Omega_{total}\approx1$ today?), monopole issue (why do not they
exist?) and finally what is the origin of the fluctuations?
\cite{7}. The inflationary scenario $(\omega<-\frac{1}{3})$ proved
to be a cornerstone to explain the mechanism of the early universe
and provides the most compelling solution of these problems. The
constant $\Lambda$ can also resolve these issues but unfortunately,
in this case inflation never ends and the normal cosmos evolution
remains impossible.

Scalar field (composed of kinetic and potential terms coupled to
gravity) is the perfect candidate to produce the dynamical framework
and acts as a source for inflation. In field theory, a spin zero
particle is usually defined by a scalar field whereas
mathematically, it is a function of space and time. It has ability
to interpret the distribution of LSS and the observed anisotropy of
the CMB radiations elegantly in inflationary era \cite{8}. Golovnev
et al. \cite{9} discussed the vector inflation models based on
orthogonal triplet and non-minimally coupled to gravity which show
the similar behavior as the scalar fields in a flat universe.

The inflation regime is divided into two parts, i.e., slow-roll and
reheating epochs. During slow-roll period, the universe inflates as
the interactions among scalar fields while other fields become
meaningless and potential energy dominates the kinetic term
\cite{10}. Reheating is the end stage of the inflation where kinetic
and potential energies are comparable and universe starts
oscillating about minimum potential \cite{11}. Warm inflation
\cite{12} has an attractive feature of joining the expanding
universe with the end of vector inflation. This motivated
researchers to discuss the inflationary scenario in the context of
warm inflation. During this regime, the dissipation effects become
strong enough due to the production of thermal fluctuations of
constant density which play a vital role in the formation of initial
fluctuations necessary for LSS construction. An additional advantage
of warm vector inflation is that the universe stops inflating and
smoothly enters into the radiation dominated phase \cite{13}.

Inflation is usually discussed in the context of intermediate and
logamediate scenarios. Intermediate inflation is motivated by string
or M theory and proved to be the exact solution of the inflationary
cosmology containing a particular form of the scale factor
\cite{14}. In intermediate era, the universe expands at the rate
slower than the standard de Sitter inflation
$(a=a_{0}\exp(H_{0}t);\quad a_{0},~H_{0}>0)$ while faster than
power-law inflation $(a=t^{m},\quad m>1)$ \cite{15}. On the other
hand, the logamediate inflation \cite{16} is motivated by applying
weak general conditions on the indefinite expanding cosmological
models. Barrow \cite{17} applied this scenario to constant,
power-law as well as exponential types of scalar field. The
effective potential and various types of potential associated with
this model are used in different DE models and in supergravity,
Kaluza-Klein as well as string theories, respectively \cite{18}. It
has been proved that the power spectrum is red or blue tilted for
this type of inflation.

In a recent paper \cite{19}, Setare and Kamali have discussed the
warm vector inflation in intermediate as well as logamediate
scenarios for FRW model and proved the physical compatibility of
these results with WMAP7 data \cite{18} through the perturbed
parameters. In this paper, we study the above mentioned scenario in
the framework of anisotropic universe model, i.e., locally
rotationally symmetric (LRS) Bianchi I (BI). The format of the paper
is as follows. In the next section, we construct the field equations
from action and calculate the corresponding pressure as well as
energy density imposing slow-roll condition. We also formulate the
slow-roll as well as scalar and tensor perturbed parameters. In
section \textbf{3}, we evaluate the Hubble parameter, scalar field
as well as slow-roll and perturbed parameters in two regimes (i)
intermediate inflation (ii) logamediate inflation. The inflationary
universe is further studied with variable and constant dissipation
factor. The results are summarized in section \textbf{4}.

\section{Basic Formalism and Perturbations}

In this section, we briefly discuss the inflationary model using the
anisotropic background of the universe. A massive vector field,
non-minimally coupled to gravity is presented by the following
action \cite{9}
\begin{equation*}
\mathcal{S}=-\frac{1}{2}\int
d^4x\sqrt{-g}(R+\frac{1}{2}F_{\alpha\beta}F^{\alpha\beta}-\frac{R}{6}
A^{\alpha}A_{\alpha}-V(A^{\alpha}A_{\alpha})),
\end{equation*}
where $\alpha,\beta=0,1,2,3$, the field strength $(F_{\alpha\beta})$
and potential $(V)$ are defined as follows
\begin{equation*}
F_{\alpha\beta}=\partial_{\alpha}A_{\beta}-\partial_{\beta}A_{\alpha},\quad
V(A^{\alpha}A_{\alpha})=m^2A^{\alpha}A_{\alpha}+\ldots
\end{equation*}
This non-minimal coupling of the vector field has an opposite
effect, i.e., it violates the conformal invariance of the massless
vector field and forces it to behave just like a minimally coupled
scalar field. The variation of the action with respect to
$A_{\alpha}$ yields the following equations of motion
\begin{equation}\label{1}
\frac{1}{\sqrt{-g}}\partial_{\alpha}(\sqrt{-g}F^{\alpha\beta})
+\frac{R}{6}A^{\beta}+\frac{\partial V}{\partial A_{\beta}}=0,
\end{equation}
where $\partial_{\alpha}=\frac{\partial}{\partial x^{\alpha}}$. An
isotropic model (FRW) is unsteady near the initial singularity thus
unable to explain the early mechanism of the universe. There is a
need of an appropriate geometry which is more concise and general
than the isotropic and homogeneous FRW universe. Bianchi models are
prime and viable models to discuss the behavior of the early
universe. A BI model being the straightforward generalization of the
flat FRW model is one of the simplest models of the anisotropic
universe.

The line element for LRS BI model $(b(t)=c(t))$ is given as
\begin{equation*}
ds^2=-dt^2+a^2(t)dx^2+b^2(t)(dy^2+dz^2),
\end{equation*}
where $a(t),~b(t)$ are the scale factors along $x$-axis and
$(y,z)$-axis, respectively. This metric can be transformed in the
following form using a linear relationship $a=b^{n},~n\neq1$
\cite{20}
\begin{equation}\label{2}
ds^2=-dt^2+b^{2n}(t)dx^2+b^2(t)(dy^2+dz^2).
\end{equation}
In this scenario, Eq.(\ref{1}) leads to
\begin{equation*}
\ddot{A}_{i}+(\frac{n+2}{3})b^{\frac{(n-4)}{3}}\dot{b}
(\dot{A}_{i}-\partial_{i}A_{0})+\frac{1}{b^{\frac{2(n+2)}{3}}}(\partial_{i}
(\partial_{k}A_{k})-\Delta
A_{i})-\partial_{i}\dot{A_{0}}+(m^2+\frac{R}{6})A_{i}=0,
\end{equation*}
\begin{equation}\label{3}
-b^{-\frac{2(n+2)}{3}}\Delta
A_{0}+(m^2+\frac{R}{6})A_{0}+b^{-\frac{2(n+2)}{3}}\partial_{i}\dot{A}_{i}=0,
\end{equation}
where dot represents the time rate of change. Applying the condition
of homogeneous vector field, i.e., $\partial_{i}A_{\alpha}=0$ to
Eq.(\ref{3}), one can deduce $A_{0}=0$. Another variable
$\mathcal{B}_{i}=\frac{A_{i}}{b^{\frac{n+2}{3}}}$ is introduced for
scalar field in place of $A_{i}$. Consequently, Eq.(\ref{3}) has the
similar form as the massive minimally coupled scalar field which can
be written as
\begin{equation*}
\ddot{\mathcal{B}}_{i}+(n+2)H_{2}\dot{\mathcal{B}}_{i}+V^{\prime}
(\mathcal{B}_{j}\mathcal{B}^{j})\mathcal{B}_{i}=0,
\end{equation*}
or equivalently
\begin{equation}\label{4}
\ddot{\mathcal{B}}_{i}+(n+2)H_{2}
\dot{\mathcal{B}}_{i}+m^2\mathcal{B}_{i}=0,
\end{equation}
where $H_{2}=\frac{\dot{b}}{b}$ is the directional Hubble parameter
and prime denotes derivative with respect to
$\mathcal{B}_{j}\mathcal{B}^{j}$.

The corresponding energy-momentum tensor can be obtained by varying
the action with respect to the metric $(g^{\alpha\beta})$ \cite{9}.
The temporal and the spatial components of energy-momentum tensor
are
\begin{eqnarray}\nonumber
T^{0}_{0}&=&\frac{1}{2}(\dot{\mathcal{B}}^{2}_{k}
+V(\mathcal{B}^{i}\mathcal{B}_{i})),\\\nonumber
T^{i}_{j}&=&\left[-\frac{5}{6}\dot{\mathcal{B}}^{2}_{k}
+\frac{1}{2}V(\mathcal{B}^2)
-\frac{2}{3}\left(\frac{n+2}{3}\right)H_{2}
\dot{\mathcal{B}}_{k}\mathcal{B}_{k}-\frac{1}{3}
\left(\left(\frac{n+2}{3}\right)\dot{H}_{2}
\right.\right.\\\nonumber&+&\left.\left.3
\left(\frac{n+2}{3}\right)^2H^2_{2}-
V^{\prime}(\mathcal{B}^2)\right)\mathcal{B}^{2}_{k}\right]
{\delta}^{i}_{j}+\dot{\mathcal{B}}_{i}\dot{\mathcal{B}}_{j}
+\left(\frac{n+2}{3}\right)H_{2}(\dot{\mathcal{B}}_{i}\mathcal{B}_{j}
+\dot{\mathcal{B}}_{j}\mathcal{B}_{i})
\\&+&\left(\left(\frac{n+2}{3}\right)\dot{H}_{2}+3
\left(\frac{n+2}{3}\right)^2H^2_{2}-
V^{\prime}(\mathcal{B}^2)\right)\mathcal{B}_{i}\mathcal{B}_{j},\label{5}
\end{eqnarray}
where $k$ stands for summation index and $\mathcal{B}_{k}$ satisfies
Eq.(\ref{4}) for any vector field. We have considered the LRS BI
universe model with homogeneous vector field which does not contain
the off-diagonal terms in the energy-momentum tensor. In order to
make the spatial component of the energy-momentum tensor diagonal,
we use various fields simultaneously. Therefore, a triplet of the
mutually orthogonal vector fields is defined as follows \cite{21}
\begin{equation*}
\sum_{a}\mathcal{B}^{a}_{i}\mathcal{B}^{a}_{j}
=|\mathcal{B}|^{2}{\delta}^{j}_{i}.
\end{equation*}
Using the above conditions with an ansatz,
$\mathcal{B}^{a}_{i}=|\mathcal{B}|{\delta}^{a}_{i}$ in Eq.(\ref{5}),
the components of the energy-momentum tensor reduce to diagonal form
as
\begin{eqnarray*}
T^{0}_{0}&=&\rho_{v}=\frac{3}{2}(\dot{\mathcal{B}}^{2}_{k}
+V(|\mathcal{B}|^{2})),\\\nonumber
T^{i}_{j}&=&-P_{v}{\delta}^{i}_{j}=
-\frac{3}{2}(\dot{\mathcal{B}}^{2}_{k}-V(|\mathcal{B}|^{2}))
{\delta}^{i}_{j}.
\end{eqnarray*}
These equations are similar to the equations which are evaluated in
\cite{22} for massive scalar field. The choice of $|\mathcal{B}|>1$
corresponds to the inflation epoch under slow-roll limit
$(\dot{\mathcal{B}}^{2}_{k}\ll V(\mathcal{B}^{2}))$ for which
$\rho_{v}=-P_{v}$.

We assume that the total energy density of the universe is the sum
of energy density of the vector field $(\rho_{v})$ and radiation
$(\rho_{\gamma})$. The mechanism of warm vector inflation can be
represented by the first field equation (evolution equation) and
conservation equations given as
\begin{eqnarray}\nonumber
H^2_{2}=\frac{1}{1+2n}(\frac{3}{2}(\dot{\mathcal{B}}^{2}_{k}
+V(|\mathcal{B}|^{2}))+\rho_{\gamma})
&=&\frac{1}{1+2n}(\rho_{v}+\rho_{\gamma}),
\\\nonumber\dot{\rho_{v}}+(n+2)H_{2}(\rho_{v}+P_{v})
&=&-\eta\dot{\mathcal{B}}^{2},\\\label{6}
\dot{\rho}_{\gamma}+\frac{4(n+2)}{3}H_{2}\rho_{\gamma}
&=&\eta\dot{\mathcal{B}}^{2},
\end{eqnarray}
where $\eta,~T_{\gamma}$ stand for dissipation or friction factor
and temperature of the thermal bath, respectively. We assume a
suitable form of $\eta=\eta_{0}\frac{T_{\gamma}^3}{\mathcal{B}^2}$,
where $\eta_{0}$ is any constant and $T_{\gamma}$ can be extracted
via quantum field theory method which holds for low temperature
\cite{23}. In this method, the inflation interacts with heavy
intermediate field (acts as catalyst) which could decay into
massless field called radiation. Here $\eta$ is taken to be positive
(by second law of thermodynamics) which implies that the energy
density of scalar field decays into radiation density. During
inflation era, $\rho_{v}\sim V(A^2)$ and the energy density of
inflation exceeds to radiation density $(\rho_{v}>\rho_{\gamma})$.
Further, we apply two limits on the above dynamic equations, i.e.,
slow-roll approximation
$(\ddot{\mathcal{B}}\ll((n+2)H_{2}+\frac{\eta}{3})\dot{\mathcal{B}})$
and quasi-stability of the radiation production where
$\dot{\rho_{\gamma}}\ll4(\frac{n+2}{3})H_{2}\rho_{\gamma},
~\dot{\rho_{\gamma}}\ll\eta\dot{\mathcal{B}}^2$ \cite{12}. Using all
these conditions, we can write Eq.(\ref{6}) as follows
\begin{eqnarray}\label{7}
-\frac{1}{2}V^{\prime}&=&(n+2)(1+\frac{\chi}{3})H_{2}\dot{\mathcal{B}},\\\label{8}
\rho_{\gamma}&=&\frac{3}{4}\chi\dot{\mathcal{B}}^{2}=\frac{(1+2n)}{8(n+2)^2}\frac{\chi}
{(1+\frac{\chi}{3})^2}\frac{{V^{\prime}}^2}{V}=C_{\gamma}T_{\gamma}^4,\\\label{9}
H^2_{2}&=&\frac{3}{2(1+2n)}V,
\end{eqnarray}
where $\chi=\frac{\eta}{(n+2)}H_{2}$ denotes the rate of dissipation
which represents strong $(\chi>1)$ and weak $(\chi<1)$ dissipation
regions. The radiation energy density is also written in the form
$C_{\gamma}T_{\gamma}^4$, where
$C_{\gamma}=\frac{\pi^2g_{\ast}}{30}$ and $g_{\ast}$ is known as the
number of relativistic degrees of freedom. Using Eqs.(\ref{7}) and
(\ref{9}) in (\ref{8}), the temperature of thermal bath can be
obtained as follows
\begin{equation}\label{10}
T_{\gamma}=\left[\frac{-(1+2n)\chi\dot{H}_{2}}{2C_{\gamma}(n+2)(1+\frac{\chi}{3})}\right]^\frac{1}{4}.
\end{equation}

The dimensionless slow-roll parameters $(\epsilon,~\lambda)$
\cite{24} of the warm vector inflation can be calculated using
Eq.(\ref{9}) as
\begin{eqnarray}\nonumber
\epsilon&=&-\frac{3}{(n+2)H_{2}}\frac{d}{dt}\left(\ln
\left(\frac{n+2}{3}H_{2}\right)\right)=-\frac{3}{(n+2)}
\frac{\dot{H}_{2}}{H^{2}_{2}}
\\\label{11}&=&\frac{1+2n}{2(n+2)^2(1+\frac{\chi}{3})}
\frac{{V^{\prime}}^2}{V^2},\\\nonumber
\lambda&=&-\frac{3}{(n+2)}\frac{\ddot{H}_{2}}{H_{2}\dot{H}_{2}}=
2\epsilon-\frac{3\dot{\epsilon}}{(n+2)H_{2}\epsilon}
\\\label{12}&=&2\epsilon+\frac{6(1+2n)}
{(n+2)^2(3+\chi)}\left(\frac{V^{\prime}}{V}\right)
\left[\frac{V^{\prime\prime}}{V^{\prime}}
-\frac{V^{\prime\prime}}{V}-\frac{3}{2}\frac{\chi^{\prime}}{(3+\chi)}\right].
\end{eqnarray}
The energy density of the scalar field can be expressed in terms of
radiation density using Eqs.(\ref{9}) and (\ref{11}) as
\begin{equation*}
\rho_{\gamma}=\frac{1}{2}\left(\frac{\chi}{3+\chi}\right)\epsilon\rho_{v}.
\end{equation*}
Here, we consider strong dissipative regime where $\chi\gg1$ which
implies that $\eta\gg3(n+2)H_{2}$. Consequently, $\rho_{\gamma}$ can
be simplified as
\begin{equation}\label{14}
\rho_{\gamma}=\frac{1}{2}\epsilon\rho_{v}.
\end{equation}
The condition of the warm inflation epoch, i.e.,
$\rho_{v}>2\rho_{\gamma}$ can be verified via an inequality
$0<\ddot{b}<1$ for which $\epsilon<1$. This inflationary scenario
ends at $\epsilon=1$ where the field energy density becomes twice of
the radiation density. The number of e-folds at two different
cosmological times $t$ (beginning of inflation) and $t_{1}$ (end of
inflation) is defined as follows
\begin{equation}\label{15}
N=\frac{(n+2)}{3}\int^{t_{1}}_{t}H_{2}dt=-\frac{(n+2)^2}{2(1+2n)}
\int^{t_{1}}_{t}(3+\chi)\frac{V}{V^{\prime}}dt.
\end{equation}

Now we evaluate scalar and tensor perturbations for anisotropic LRS
BI universe model at small scale structure by varying the field
$\mathcal{B}$. In non-warm and warm inflation, quantum and thermal
fluctuations, respectively yield \cite{25}
\begin{equation}\label{16}
\langle\delta
\mathcal{B}\rangle_{quantum}=\left(\frac{n+2}{3}\right)^2\frac{H^{2}_{2}}{2\pi},
\quad\langle\delta
\mathcal{B}\rangle_{thermal}=\left(\frac{n+2}{3(4\pi)^3}\right)^\frac{1}{4}(\eta
H_{2}T_{\gamma}^2)^\frac{1}{4}.
\end{equation}
In cosmology, a useful function of wave number $(k)$, known as power
spectrum is introduced to quantify the variance in the fluctuations
due to inflation. In warm inflationary universe, the scalar power
spectrum is given by \cite{26}
\begin{equation}\label{17}
\Delta^{2}_{R}(k)=\left(\frac{n+2}{3}\right)^2\left(\frac{H_{2}}
{\dot{\mathcal{B}}}\langle\delta
\mathcal{B}\rangle_{thermal}\right)^2.
\end{equation}
Using Eqs.(\ref{6}), (\ref{9}) and (\ref{16}), we can calculate the
power-spectrum of scalar perturbation in the form
\begin{equation}\label{18}
\Delta^{2}_{R}(k)=-\left[\frac{(n+2)^5\eta^3T_{\gamma}^2}{972(1+2n)^2(4\pi)^3}\right]
^\frac{1}{2}\frac{H^\frac{3}{2}_{2}}{\dot{H}_{2}}=\left[\frac{(n+2)^5\eta^5T_{\gamma}^2}
{6^{\frac{3}{2}}972(1+2n)^\frac{5}{2}(4\pi)^3}\right]
^\frac{1}{2}\frac{V^\frac{5}{4}}{{V^{\prime}}^2}.
\end{equation}
The second important parameter to study the fluctuations is the
scalar spectral index $(n_{s})$. For the present model, it is
defined as
\begin{equation}\label{19}
n_{s}-1=-\frac{d\ln\Delta^{2}_{R}(k)}{d\ln k}.
\end{equation}
The tensor perturbation $(\Delta^{2}_{T}(k))$ and its spectral index
$(n_{T})$ for anisotropic universe are
\begin{eqnarray}\label{20}
\Delta^{2}_{T}(k)&=&\left(\frac{n+2}{3\pi}\right)^2H^{2}_{2}=
\left(\frac{(n+2)^2}{3(1+2n)}\right)\frac{V}{\pi^2},\\\label{21a}
n_{T}&=&-2\epsilon.
\end{eqnarray}
The tensor-scalar ratio $(R)$ has the following form
\begin{equation}\label{21}
R=-\left[\frac{48(1+2n)(4\pi)^3}{(n+2)\pi^4\eta^3T_{\gamma}^2}\right]^\frac{1}{2}
H^{\frac{1}{2}}_{2}\dot{H}_{2}.
\end{equation}
According to the observations of WMAP+BAO+SN, the perturbed scalar
power spectrum is constrained to
$\Delta^{2}_{R}(k_{0}=0.002Mpc^{-1})=(2.445\pm0.096)\times10^{-9}$
\cite{8}. In this context, the physical acceptable range of
scalar-tensor ratio is determined, i.e., $R<0.22$ representing the
expanding universe.

\section{Intermediate and Logamediate Inflation}

In this section, we discuss the warm vector inflation in the context
of intermediate and logamediate inflations by treating $\eta$ as a
function of $\mathcal{B}$ as well as a constant.

\subsection{Intermediate Inflation}

In intermediate scenario, the scale factor follows the law \cite{14}
\begin{equation}\label{22}
b(t)=b_{0}\exp(\mu t^{g}), \quad \mu>0,~0<g<1.
\end{equation}
The number of e-folds for this model can be calculated using
Eq.(\ref{15}) as
\begin{equation}\label{23}
N=\frac{(n+2)}{3}\mu(t^{g}-t^{g}_{1}).
\end{equation}
Firstly, we explore the dynamics of the warm intermediate inflation
by considering the variable dissipation factor.

\subsubsection{Case 1: $\eta=\eta_{0}\frac{T_{\gamma}^3}{\mathcal{B}^2},
~\eta_{0}=constant$}

Here, we evaluate some quantities which are useful to determine the
parameters defined in the previous section. The scalar potential is
evaluated in terms of cosmic time $t$ using Eq.(\ref{9}) as
\begin{equation}\label{24}
V=\frac{2(1+2n)}{3}(\mu g)^2t^{2(g-1)}.
\end{equation}
The scalar field and the directional Hubble parameter are found
using Eqs.(\ref{7}), (\ref{9}), (\ref{10}) and (\ref{24}) as follows
\begin{equation}\label{25}
\mathcal{B}=\mathcal{B}_{0}\exp(\omega_{0}t^\frac{5g+2}{8}),\quad
H_{2}=\mu g\left(\frac{\ln \mathcal{B}-\ln
\mathcal{B}_{0}}{\omega_{0}}\right)^\frac{8(g-1)}{5g+2},
\end{equation}
where
$\omega_{0}=[\frac{2}{3\eta_{0}}(\frac{1+2n}{n+2})^\frac{1}{4}(\frac{2C_{\gamma}}{3})
^\frac{3}{4}]^\frac{1}{2}\frac{(1-g)^\frac{1}{8}(\mu
g)^\frac{5}{8}}{5g+2}$ is a pure constant. In this context, the
slow-roll parameters take the form
\begin{eqnarray}\nonumber
\epsilon&=&\left(\frac{3}{n+2}\right)\frac{1-g}{\mu
g}\left(\frac{\ln \mathcal{B}-\ln
\mathcal{B}_{0}}{\omega_{0}}\right)^\frac{-8g}{5g+2},\\\label{26}
\lambda&=&\left(\frac{3}{n+2}\right)\frac{2-g}{\mu g}\left(\frac{\ln
\mathcal{B}-\ln
\mathcal{B}_{0}}{\omega_{0}}\right)^\frac{-8g}{5g+2}.
\end{eqnarray}
The radiation density (\ref{14}) can be calculated in terms of
scalar field as
\begin{equation}\label{27}
\rho_{\gamma}=\frac{3(1+2n)}{2}\epsilon
H^{2}_{2}=\frac{9(1+2n)}{2(n+2)}(\mu g)(1-g)\left(\frac{\ln
\mathcal{B}-\ln
\mathcal{B}_{0}}{\omega_{0}}\right)^\frac{8(g-2)}{5g+2}.
\end{equation}
Inflationary era $(\epsilon<1)$ has the scalar field of magnitude
$\mathcal{B}$.

We assume that $\mathcal{B}_{1}$ is another field which is produced
at the end of inflation and could be found by fixing $\epsilon=1$ as
\begin{equation*}
\mathcal{B}_{1}=\mathcal{B}_{0}\exp\left[\omega_{0}\left(\frac{3(1-g)}{(m+2)\mu
g}\right) ^\frac{5g+2}{8g}\right].
\end{equation*}
Inserting the values of two different cosmic times (using
$\mathcal{B}$ and $\mathcal{B}_{1}$) in Eq.(\ref{23}), we get
\begin{equation*}
N=(\frac{n+2}{3})\mu\left[\left(\frac{\ln \mathcal{B}-\ln
\mathcal{B}_{0}}{\omega_{0}}\right)^\frac{8(g-2)}{5g+2}-\left(\frac{\ln
\mathcal{B}_{1}-\ln
\mathcal{B}_{0}}{\omega_{0}}\right)^\frac{8(g-2)}{5g+2}\right].
\end{equation*}
These two equations yield $\mathcal{B}$ in terms of $N$ in the
following form
\begin{equation}\label{30}
\mathcal{B}=\mathcal{B}_{0}\exp\left[\omega_{0}\left(\left(\frac{3}{n+2}\right)\left(\frac{N}{\mu}
+\frac{1-g}{g\mu}\right) \right)^\frac{5g+2}{8g}\right].
\end{equation}
Using Eqs.(\ref{10}) and (\ref{25}) in (\ref{18}), the perturbed
scalar power spectrum in warm vector intermediate inflation (can
also be expressed in terms of e-folds with the help of
Eq.(\ref{30})) is
\begin{eqnarray}\nonumber
\Delta^{2}_{R}(k)&=&\left(\frac{\eta^{3}_{0}}{972(4\pi)^3}\right)^{\frac{1}{2}}
\left[\frac{3^{11}(n+2)^{29}(\mu
g)^{15}(1-g)^3}{(2C_{\gamma})^{11}(1+2n)^5}\right]
^\frac{1}{8}\mathcal{B}^{-3}\\\nonumber&\times&\left(\frac{\ln
\mathcal{B}-\ln
\mathcal{B}_{0}}{\omega_{0}}\right)^\frac{15g-18}{5g+2}\\\nonumber&=&
\left(\frac{\eta^{3}_{0}}{972(4\pi)^3}\right)
^{\frac{1}{2}}\left[\frac{3^{11}(n+2)^{29}(\mu
g)^{15}(1-g)^3}{(2C_{\gamma})^{11}(1+2n)^5}\right]
^\frac{1}{8}\mathcal{B}^{-3}_{0}\\\nonumber
&\times&\exp\left[-3\omega_{0}\left(\left(\frac{3}{m+2}\right)\left(\frac{N}{\mu}
+\frac{1-g}{g\mu}\right)\right)^\frac{5g+2}{8g}\right]
\\\nonumber&\times&\left[\left(\frac{3}{n+2}\right)\left(\frac{N}{\mu}
+\frac{1-g}{g\mu}\right)\right]^\frac{15g-18}{8g}.
\end{eqnarray}
Using this equation in the scalar spectral index (\ref{19}), we have
\begin{eqnarray}\nonumber
n_{s}&=&1+\left(\frac{3}{n+2}\right)\left(\frac{15g-18}{8g\mu}\right)\left(\frac{\ln
\mathcal{B}-\ln \mathcal{B}_{0}}{\omega_{0}}\right)^\frac{-8g}{5g+2}
\\\label{31}&=&1+\left(\frac{3}{n+2}\right)\left(\frac{15g-18}{8g\mu}\right)
\left[\left(\frac{3}{n+2}\right)\left(\frac{N}{\mu}
+\frac{1-g}{g\mu}\right)\right]^{-1}.
\end{eqnarray}
\begin{figure}
\center\epsfig{file=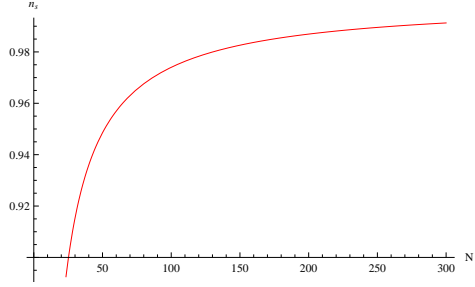, width=0.45\linewidth}\caption{The
graph of $n_{s}$ versus number of e-folds for
$\mu=1,~g=\frac{1}{2},~n\approx0.5$ in intermediate scenario.}
\end{figure}

Figure \textbf{1} shows the increasing behavior of $n_{s}$ with
respect to $N$. The observational value of $n_{s}=0.96$ corresponds
to $N=60$ which indicates the physical compatibility of this
anisotropic model with WMAP7 data. Equations (\ref{20}) and
(\ref{21a}) give the tensor power spectrum as well as spectral
index, respectively as follows
\begin{eqnarray}\nonumber
\Delta^{2}_{T}(k)&=&\frac{2}{9\pi^2}(n+2)^2(\mu g)^2\left(\frac{\ln
\mathcal{B}-\ln
\mathcal{B}_{0}}{\omega_{0}}\right)^\frac{16(g-1)}{5g+2}
\\\nonumber&=&\frac{2}{9\pi^2}(n+2)^2(\mu g)^2
\left[\left(\frac{3}{n+2}\right)\left(\frac{N}{\mu}
+\frac{1-g}{g\mu}\right)\right]^{\frac{2(g-1)}{g}},\\\nonumber
n_{T}&=&-2\left(\frac{3}{n+2}\right)\left(\frac{1-g}{\mu
g}\right)\left(\frac{\ln \mathcal{B}-\ln
\mathcal{B}_{0}}{\omega_{0}}\right)^\frac{-8g}{5g+2}\\\nonumber
&=&-\frac{6(1-g)}{(n+2)\mu
g}\left[\left(\frac{3}{m+2}\right)\left(\frac{N}{\mu}
+\frac{1-g}{g\mu}\right)\right]^{-1}.
\end{eqnarray}
The tensor-scalar ratio is obtained as
\begin{eqnarray}\nonumber
R&=&\left(\frac{48(4\pi)^3(n+2)^4(\mu
g)^4}{\pi^4\eta^{3}_{0}}\right)
^{\frac{1}{2}}\left[\frac{(2C_{\gamma})^{11}(1+2n)^5}{3^{11}(n+2)^{29}(\mu
g)^{15}(1-g)^3}\right]
^\frac{1}{8}\\\nonumber&\times&\mathcal{B}^{-3}\left(\frac{\ln
\mathcal{B}-\ln
\mathcal{B}_{0}}{\omega_{0}}\right)^\frac{g+2}{5g+2}\\\nonumber&=&
\left(\frac{48(4\pi)^3(n+2)^4(\mu g)^4}{\pi^4\eta^{3}_{0}}\right)
^{\frac{1}{2}}\left[\frac{(2C_{\gamma})^{11}(1+2n)^5}{3^{11}(n+2)^{29}(\mu
g)^{15}(1-g)^3}\right] ^\frac{1}{8}\mathcal{B}^{3}_{0}\\\nonumber
&\times&\exp\left[3\omega_{0}\left(\left(\frac{3}{n+2}\right)\left(\frac{N}{\mu}
+\frac{1-g}{g\mu}\right)\right)^\frac{5g+2}{8g}\right]
\\\nonumber
&\times&\left[\left(\frac{3}{n+2}\right)\left(\frac{N}{\mu}
+\frac{1-g}{g\mu}\right)\right]^\frac{g+2}{8g}.
\end{eqnarray}
Rewriting the above equation in the form of $n_{s}$, we have
\begin{eqnarray}\nonumber
R&=&\left(\frac{48(4\pi)^3(n+2)^4(\mu
g)^4}{\pi^4\eta^{3}_{0}}\right)
^{\frac{1}{2}}\left[\frac{(2C_{\gamma})^{11}(1+2n)^5}{3^{11}(n+2)^{29}(\mu
g)^{15}(1-g)^3}\right]
^\frac{1}{8}\mathcal{B}^{3}_{0}\\\nonumber&\times&\exp\left[3\omega_{0}\left(\left(\frac{3}{n+2}\right)
\left(\frac{18-15g}{8g\mu(1-n_{s})}\right)\right)^\frac{5g+2}{8g}\right]
\\\label{34}
&\times&\left[\left(\frac{3}{n+2}\right)
\left(\frac{18-15g}{8g\mu(1-n_{s})}\right)\right]^\frac{g+2}{8g}.
\end{eqnarray}
The left graph of Figure \textbf{2} shows that the anisotropic model
is incompatible with WMAP7 data for all the three choices of
dissipation factor. With the help of fine-tuning of the parameters
$g$ and $\mu$, we are able to find the range $R<0.22$ in which
$n_{s}=0.96$ lies for $\eta_{0}=0.25,~1,~4$. The compatible behavior
of the tensor-scalar ratio with spectral index is shown in the right
graph.
\begin{figure}
\center\epsfig{file=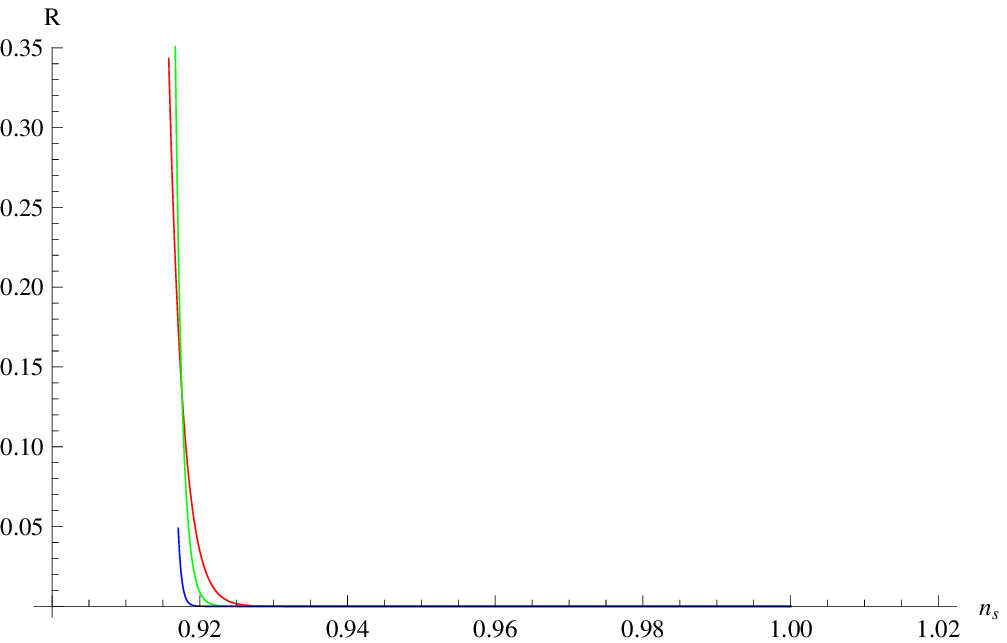,
width=0.45\linewidth}\epsfig{file=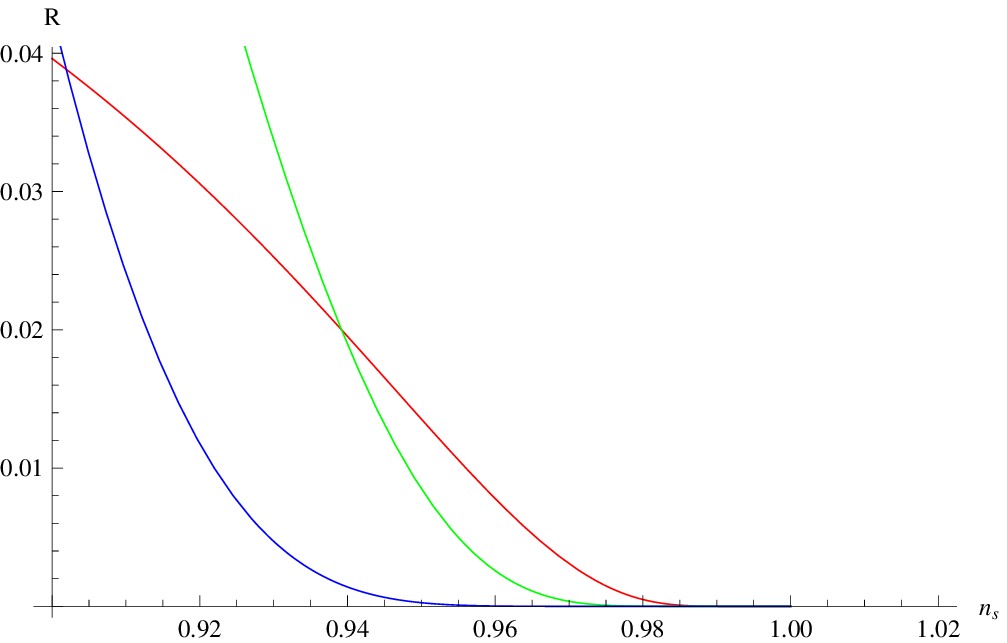,
width=0.45\linewidth}\caption{The left graph of scalar-tensor ratio
versus $n_{s}$ for
$\mu=1,~C_{\gamma}=70,~g=\frac{1}{2},~\mathcal{B}_{0}\propto
C_{\gamma}^{-\frac{1}{12}},~n\approx0.5,~\eta_{0}=0.25$(blue),
$1$(green), $4$(red). The right graph is plotted for
$\mu=5,~g=0.95$.}
\end{figure}

\subsubsection{Case 2: $\eta=\eta_{1}=constant$}

Now, we calculate all the above parameters by taking the constant
dissipation factor. When $\eta$ is constant, $\mathcal{B}$ and
$H_{2}$ take the following form
\begin{equation}\label{35}
\mathcal{B}=\mathcal{B}_{0}+\omega_{1}t^\frac{2g-1}{2},\quad
H_{2}=\mu g\left(\frac{\mathcal{B}-
\mathcal{B}_{0}}{\omega_{1}}\right)^\frac{2(g-1)}{2g-1},
\end{equation}
with $\omega_{1}=\left[\frac{8(1+2n)}{\eta_{1}(2g-1)^2}(1-g)(\mu
g)^2\right]^\frac{1}{2}$. The corresponding slow-roll parameters are
\begin{equation}\label{36}
\epsilon=(\frac{3}{n+2})\frac{1-g}{\mu g}\left(\frac{\mathcal{B}-
\mathcal{B}_{0}}{\omega_{1}}\right)^\frac{2g}{1-2g},\quad
\lambda=(\frac{3}{n+2})\frac{2-g}{\mu g}\left(\frac{\mathcal{B}-
\mathcal{B}_{0}}{\omega_{1}}\right)^\frac{2g}{1-2g}.
\end{equation}
In this case, the relationship between $\rho_{\gamma}$ and
$\rho_{v}$ becomes
\begin{equation*}
\rho_{\gamma}=\frac{9(1+2n)}{2(n+2)}(\mu
g)(1-g)\left(\frac{\mathcal{B}-
\mathcal{B}_{0}}{\omega_{1}}\right)^\frac{4-2g}{1-2g}.
\end{equation*}
The number of e-folds between $\mathcal{B}$ and $\mathcal{B}_{1}$
are calculated as
\begin{equation}\label{37}
N=(\frac{m+2}{3})\mu\left[\left(\frac{\mathcal{B}-
\mathcal{B}_{0}}{\omega_{1}}\right)^\frac{2g}{2g-1}-\left(\frac{\mathcal{B}_{1}-
\mathcal{B}_{0}}{\omega_{1}}\right)^\frac{2g}{2g-1}\right],
\end{equation}
where
\begin{equation*}
\mathcal{B}_{1}=\mathcal{B}_{0}+\omega_{1}\left[\frac{3(1-g)}{(n+2)\mu
g}\right]^\frac{5g+2}{8g}.
\end{equation*}
The scalar field in terms of e-folds is given as
\begin{equation}\label{38}
\mathcal{B}=\mathcal{B}_{0}+\omega_{1}\left[\left(\frac{3}{n+2}\right)\left(\frac{N}{\mu}
+\frac{1-g}{g\mu}\right)\right]^\frac{2g-1}{2g}.
\end{equation}
\begin{figure}
\center\epsfig{file=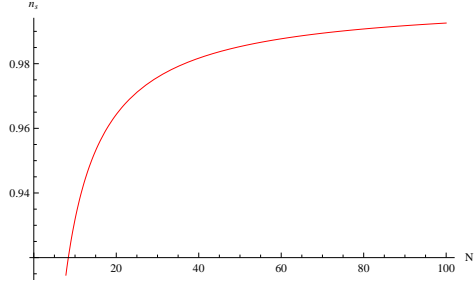, width=0.45\linewidth}\caption{The
graph of $n_{s}$ versus number of e-folds.}
\end{figure}

The corresponding scalar power spectrum and spectral index are
\begin{eqnarray}\nonumber
\Delta^{2}_{R}(k)&=&
\left[\left(\frac{3}{2C_{\gamma}}\right)^\frac{1}{2}
\frac{(n+2)^\frac{9}{2}\eta^{3}_{1}(\mu g)^{\frac{3}{2}}}
{972(4\pi)^3(1+2n)^\frac{3}{2}(1-g)^\frac{3}{2}}\right]^{\frac{1}{2}}
\\\label{39}&\times&\left[\left(\frac{3}{n+2}\right)\left(\frac{N}{\mu}
+\frac{1-g}{g\mu}\right)\right]^{\frac{3}{4}},\\\nonumber
n_{s}&=&1-\left(\frac{9}{4\mu(n+2)}\right)
\left[\left(\frac{3}{n+2}\right)\left(\frac{N}{\mu}
+\frac{1-g}{g\mu}\right)\right]^{-1}.
\end{eqnarray}
The graphical behavior of the spectral index against number of
e-folds is shown in Figure \textbf{3}. The corresponding number $N$
of $n_{s}=0.96$ decreases as compared to the previous case in which
$\eta$ is a function of $\mathcal{B}$. In this case, the model
remains consistent with WMAP7 observations. Similarly,
$\Delta^{2}_{T}$ and $n_{T}$ can be written as
\begin{eqnarray}\label{41}
\Delta^{2}_{T}&=&\frac{2}{9\pi^2}(n+2)^2(\mu g)^2
\left[\left(\frac{3}{n+2}\right)\left(\frac{N}{\mu}
+\frac{1-g}{g\mu}\right)\right]^{\frac{2(g-1)}{g}},\\\nonumber
n_{T}&=&-\frac{6(1-g)}{\mu
g(m+2)}\left[\left(\frac{3}{n+2}\right)\left(\frac{N}{\mu}
+\frac{1-g}{g\mu}\right)\right]^{-1}.
\end{eqnarray}
The tensor-scalar ratio can be found as
\begin{eqnarray}\nonumber
R&=&\left[\frac{48(4\pi)^3(1+2n)^\frac{3}{2}(\mu g)^\frac{5}{2}(1-g)
^\frac{3}{2}(2C_{\gamma})^{\frac{1}{2}}}
{\pi^4\eta^{3}_{1}3^{\frac{1}{2}}(n+2)^\frac{1}{2}}\right]^{\frac{1}{2}}
\\\nonumber&\times&\left[\left(\frac{3}{n+2}\right)\left(\frac{N}{\mu}
+\frac{1-g}{g\mu}\right)\right]^{\frac{5g-8}{4g}},\\\nonumber&=&
\left[\frac{48(4\pi)^3(1+2n)^\frac{3}{2}(\mu g)^\frac{5}{2}(1-f)
^\frac{3}{2}(2C_{\gamma})^{\frac{1}{2}}}
{\pi^4\eta^{3}_{1}3^{\frac{1}{2}}(n+2)^\frac{1}{2}}\right]^{\frac{1}{2}}
\left[\frac{4\mu(n+2)}{9}(1-n_{s})\right]^{\frac{8-5g}{4g}}.\\\label{42}
\end{eqnarray}
Figure \textbf{4} shows the agreement of the considered model with
WMAP data as the value of observational interest of $n_{s}$ lies in
the region $R<0.22$.
\begin{figure}
\center\epsfig{file=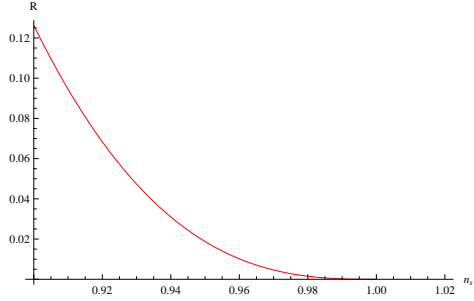, width=0.45\linewidth}\caption{The
graph of scalar-tensor ratio versus $n_{s}$ for
$\mu=1,~C_{\gamma}=70,~g=\frac{1}{2},~n\approx0.5,~\eta_{1}\propto
C_{\gamma}^{\frac{1}{6}}$.}
\end{figure}

\subsection{Logamediate Inflation}

Here we evaluate the above mentioned parameters in the context of
logamediate inflation to study the dynamics of warm vector
inflation. In this scenario, the scale factor has the specific form
\cite{16}
\begin{equation}\label{43}
b(t)=b_{0}\exp(\mu[\ln t]^{\psi}),\quad\psi>1,~\mu>0.
\end{equation}
Thus the number of e-folds is
\begin{equation}\label{44}
N=\frac{(n+2)}{3}\mu([\ln t]^{\psi}-[\ln t_{1}]^{\psi}).
\end{equation}

\subsubsection{Case 1: $\eta=\eta_{0}\frac{T_{\gamma}^3}{\mathcal{B}^2}$}

In this case, the scalar field and the directional Hubble parameter
are
\begin{eqnarray}\label{45}
\mathcal{B}&=&\mathcal{B}_{0}\exp(\omega_{2}\Gamma(t)),\\\nonumber
H_{2}&=&\mu\psi\left(\ln\left[\Gamma^{-1}\left(\frac{\ln
\mathcal{B}-\ln
\mathcal{B}_{0}}{\omega_{2}}\right)\right]\right)^{2(\psi-1)}\left[\Gamma^{-1}\left(\frac{\ln
\mathcal{B}-\ln
\mathcal{B}_{0}}{\omega_{2}}\right)\right]^{-2},\\\label{46}
\end{eqnarray}
where
$\omega_{2}=\left[\frac{2(2C_{\gamma})^{\frac{3}{4}}(\mu\psi)^{\frac{5}{4}}(n+2)
^\frac{3}{4}(1+2n)^{\frac{1}{4}}}{3^{\frac{7}{4}}\eta_{0}}\right]^{\frac{1}{2}}$
and $\Gamma(t)=\gamma[\frac{5\psi+3}{8},\frac{\ln t}{4}]$ is the
incomplete gamma function. The potential term is expressed in terms
of $\mathcal{B}$ using Eqs.(\ref{9}) and (\ref{45}) as
\begin{eqnarray}\nonumber
V&=&\frac{2(1+2n)}{3}(\mu\psi)^2\left(\ln\left[\Gamma^{-1}\left(\frac{\ln
\mathcal{B}-\ln
\mathcal{B}_{0}}{\omega_{2}}\right)\right]\right)^{2(\psi-1)}\\\label{47}
&\times&\left[\Gamma^{-1}\left(\frac{\ln \mathcal{B}-\ln
\mathcal{B}_{0}}{\omega_{2}}\right)\right]^{-2}.
\end{eqnarray}
The slow-roll parameters can be described as
\begin{eqnarray}\nonumber
\epsilon&=&\frac{3}{(n+2)\mu\psi}\left(\ln\left[\Gamma^{-1}\left(\frac{\ln
\mathcal{B}-\ln
\mathcal{B}_{0}}{\omega_{2}}\right)\right]\right)^{2(\psi-1)},\\\label{49}
\lambda&=&\frac{6}{(n+2)\mu\psi}\left(\ln\left[\Gamma^{-1}\left(\frac{\ln
\mathcal{B}-\ln
\mathcal{B}_{0}}{\omega_{2}}\right)\right]\right)^{2(\psi-1)}.
\end{eqnarray}
The corresponding energy density of radiation is
\begin{eqnarray*}
\rho_{\gamma}&=&\frac{9(1+2n)}{2(m+2)}(\mu\psi)
\left(\ln\left[\Gamma^{-1}\left(\frac{\ln \mathcal{B}-\ln
\mathcal{B}_{0}}{\omega_{2}}\right)\right]\right)^{6(\psi-1)}
\\\nonumber&\times&\left(\Gamma^{-1}\left(\frac{\ln
\mathcal{B}-\ln \mathcal{B}_{0}}{\omega_{2}}\right]\right)^{-4}.
\end{eqnarray*}

The number of e-folds between two fields is given as
\begin{equation}\label{50}
N=\left(\frac{n+2}{3}\right)\mu\left[\left(\ln\left[\Gamma^{-1}\left(\frac{\ln
\mathcal{B}-\ln
\mathcal{B}_{0}}{\omega_{2}}\right)\right]\right)^{\psi}
-\left(\frac{(n+2)}{3}\mu\psi\right) ^{\frac{\psi}{1-\psi}}\right].
\end{equation}
The second equality in the above equation is obtained by using the
value of $\mathcal{B}_{1}$ and putting $\epsilon=1$. Equation
(\ref{45}) can be rewritten in terms of e-folds as
\begin{equation}\label{51}
\mathcal{B}=\mathcal{B}_{0}\exp\left[\omega_{2}\Gamma\exp\left[\left(\frac{3}{n+2}\right)\frac{N}{\mu}
+\left(\frac{3}{n+2}\mu\psi\right)
^\frac{\psi}{1-\psi}\right]^{\frac{1}{\psi}}\right].
\end{equation}
The corresponding perturbed parameters $\Delta^{2}_{R}(k)$ and
$\Delta^{2}_{T}(k)$ are
\begin{eqnarray}\nonumber
\Delta^{2}_{R}(k)&=&\left(\frac{\eta^{3}_{0}}{(4\pi)^3}\right)^{\frac{1}{2}}
\left[\frac{3^{11}(m+2)^{9}(\mu\psi)^{15}(1+2n)^3}{(2C_{\gamma})^{11}}\right]
^\frac{1}{8}\mathcal{B}^{-3}_{0}\\\nonumber&\times&\exp
\left[-3\omega_{2}\Gamma\exp\left[\left(\frac{3}{n+2}\right)\frac{N}{\mu}
+\left(\frac{3}{n+2}(\mu\psi)\right)
^\frac{\psi}{1-\psi}\right]^{\frac{1}{\psi}}\right]
\\\nonumber&\times&\exp\left[\frac{-15}{8}\left(\exp\left[\left(\frac{3}{n+2}\right)\frac{N}{\mu}
+\left(\frac{3}{n+2}(\mu\psi)\right)
^\frac{\psi}{1-\psi}\right]^{\frac{1}{\psi}}\right)\right]
\\\label{52}&\times&\left[\left(\frac{3}{n+2}\right)\frac{N}{\mu}
+\left(\frac{3}{n+2}(\mu\psi)\right)
^\frac{\psi}{1-\psi}\right]^{\frac{15(\psi-1)}{8\psi}},\\\nonumber
\Delta^{2}_{T}&=&\frac{2}{9\pi^2}(m+2)^2(\mu\psi)^2\left[\left(\frac{3}{n+2}\right)\frac{N}{\mu}
+\left(\frac{3}{n+2}(\mu\psi)\right)
^\frac{\psi}{1-\psi}\right]^{\frac{2(\psi-1)}{\psi}}
\\\label{53}&\times&\exp\left[-2\left(\left(\frac{3}{n+2}\right)\frac{N}{\mu}
+\left(\frac{3}{n+2}(\mu\psi)\right)
^\frac{\psi}{1-\psi}\right)^{\frac{1}{\psi}}\right].
\end{eqnarray}
The corresponding spectral indices are
\begin{eqnarray}\nonumber
n_{s}-1&=&\frac{15(1-\psi)}{8\mu\psi}\left(\frac{3}{n+2}\right)
\left[\left(\frac{3}{n+2}\right)\frac{N}{\mu}
+\left(\frac{3}{n+2}(\mu\psi)\right)
^\frac{\psi}{1-\psi}\right]^{-1},\\\label{54}\\\nonumber
n_{T}&=&-\frac{6}{(n+2)\mu\psi}\left[\left(\frac{3}{n+2}\right)\frac{N}{\mu}
+\left(\frac{3}{n+2}(\mu\psi)\right)
^\frac{\psi}{1-\psi}\right]^{1-\psi}.
\end{eqnarray}
\begin{figure}
\center\epsfig{file=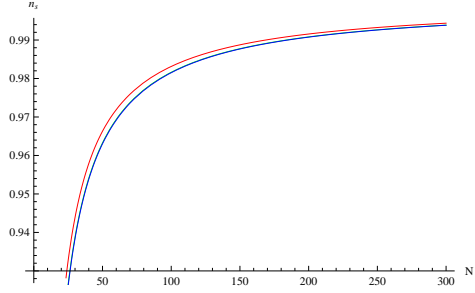, width=0.45\linewidth}\caption{The
graph of $n_{s}$ versus number of e-folds for $\mu=1,~\psi=10$(red),
$50$(blue), $n\approx0.5$ in logamediate scenario.}
\end{figure}

Figure \textbf{5} indicates the behavior of $n_{s}$ against $N$ for
$\psi=10,~50$. The number of e-folds decreases as $\psi$ increases.
Here the allowed value of the parameter $n_{s}=0.96$ lies in the
range $N\approx50$. Finally, we obtain the following observational
parameter of interest as a function of spectral index
\begin{eqnarray}\nonumber
R&=&\left(\frac{4(4\pi)^3}{81\pi^4\eta^{3}_{0}}
\right)^{\frac{1}{2}}
\left[\frac{(2C_{\gamma})^{11}(n+2)^{7}(\mu\psi)}{3^{11}(1+2n)^3}\right]
^\frac{1}{8}\mathcal{B}^{3}_{0}\left[\left(\frac{3}{n+2}\right)
\frac{15(\psi-1)}{8\mu\psi(1-n_{s})}\right]
^{\frac{1}{\psi}}\\\nonumber&\times&\exp
\left[3\omega_{2}\Gamma\exp\left[\left(\frac{3}{n+2}\right)
\frac{15(\psi-1)}{8\mu\psi(1-n_{s})}\right]
^{\frac{1}{\psi}}\right]\\\label{56}&\times&\exp
\left[-\frac{1}{8}\left[\left(\frac{3}{n+2}\right)
\frac{15(\psi-1)}{8\mu\psi(1-n_{s})}
\right]^{\frac{1}{\psi}}\right].
\end{eqnarray}
We see from Figure \textbf{6} the consistency of the model with
observational data for specific values of $\eta_{0}$. The ratio $R$
decreases as $\psi$ increases.
\begin{figure}
\center\epsfig{file=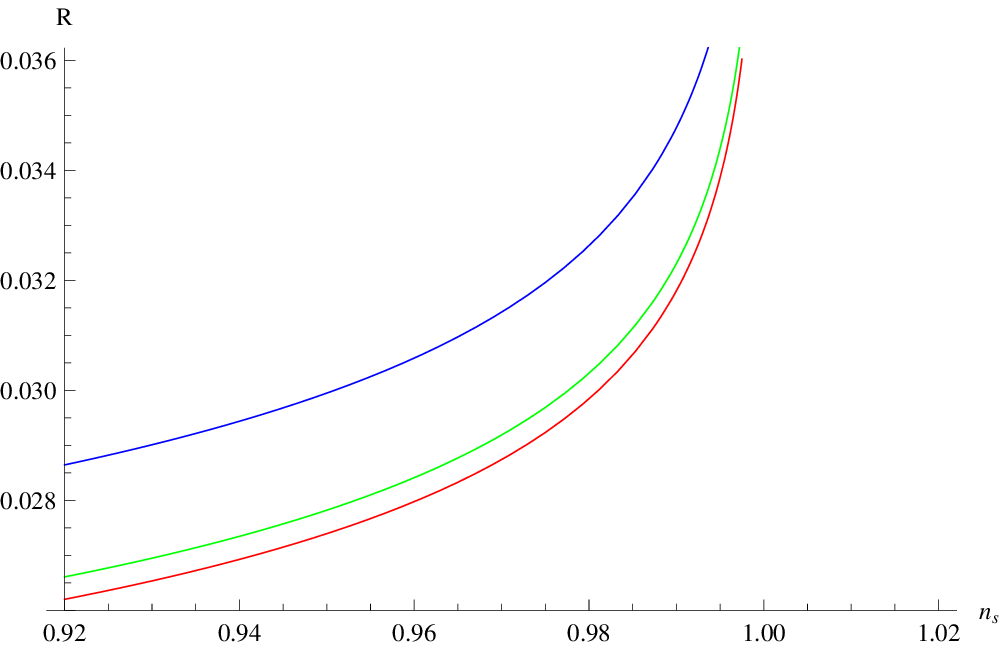,
width=0.45\linewidth}\epsfig{file=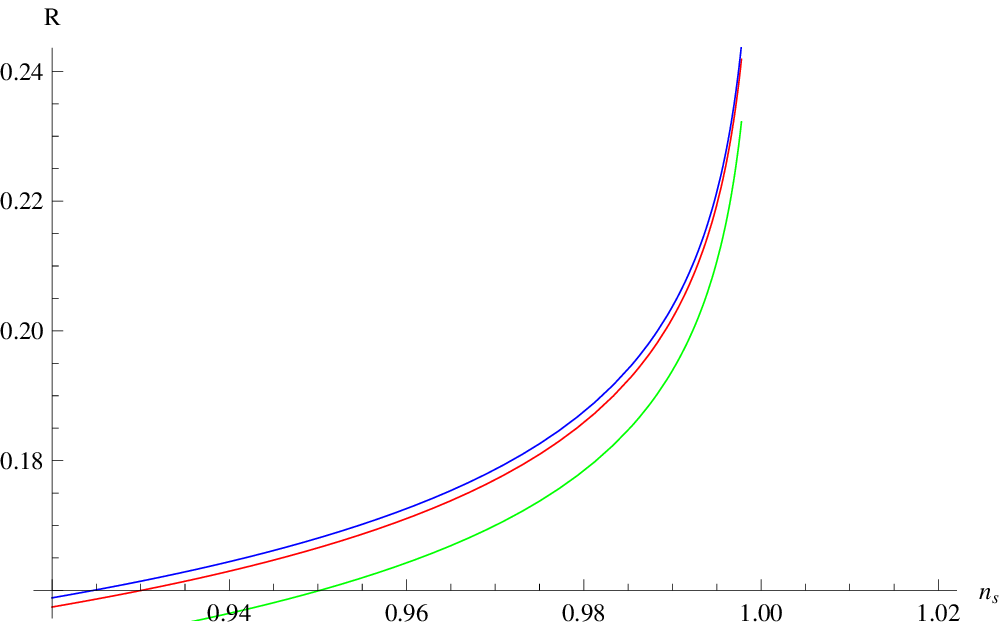,
width=0.45\linewidth}\caption{The left graph of scalar-tensor ratio
versus $n_{s}$ for
$\mu=1,~C_{\gamma}=70,~f=\frac{1}{2},~n\approx0.5,~\mathcal{B}_{0}\propto
C_{\gamma}^{-\frac{1}{12}},~\psi=10,~\eta_{0}=0.25$(blue),
$1$(green), $4$(red). The right graph is plotted for $\psi=50$.}
\end{figure}

\subsection{Case 2: $\eta=\eta_{1}$}

For this case, the scalar field and the directional Hubble parameter
are
\begin{equation}\label{57}
\mathcal{B}=\mathcal{B}_{0}+\omega_{3}\Gamma(t),\quad
H_{2}=\frac{\mu\psi\left(\ln\left[\Gamma^{-1}\left(\frac{\mathcal{B}-
\mathcal{B}_{0}}{{\omega}_{3}}\right)\right]\right)^{(\psi-1)}}
{\Gamma^{-1}\left(\frac{
\mathcal{B}-\mathcal{B}_{0}}{{\omega}_{3}}\right)},
\end{equation}
where ${\omega}_{3}=\left[\frac{2(1+2n)(\mu\psi)^{2}}
{\eta_{1}}\right]^{\frac{1}{2}} \frac{(\psi-1)!}{2^{\psi-1}}$ and
$\Gamma(t)=\gamma[\psi,\frac{\ln t}{2}]$. In this case, $V$
transformed into
\begin{equation*}
V=\frac{2(1+2n)}{3}\left[\frac{(\mu\psi)\left(\ln\left[\Gamma^{-1}\left(\frac{\mathcal{B}-
\mathcal{B}_{0}}{\omega_{3}}\right)\right]\right)^{(\psi-1)}}{{\Gamma^{-1}\left(\frac{
\mathcal{B}-\mathcal{B}_{0}}{\omega_{3}}\right)}}\right]^2.
\end{equation*}
The parameters $\epsilon$ and $\lambda$ take the following form
\begin{equation}\label{59}
\epsilon=\frac{3\left(\ln\left[\Gamma^{-1}\left(\frac{\mathcal{B}-
\mathcal{B}_{0}}{\omega_{3}}\right)\right]\right)^{1-\psi}}{(n+2)(\mu\psi)},\quad
\lambda=\frac{6\left(\ln\left[\Gamma^{-1}\left(\frac{\mathcal{B}-
\mathcal{B}_{0}}{\omega_{3}}\right)\right]\right)^{1-\psi}}{(n+2)(\mu\psi)}.
\end{equation}
The number of e-folds becomes
\begin{eqnarray}\label{60}
N=\left(\frac{n+2}{3}\right)\mu\left[\left(\ln\left[\Gamma^{-1}\left(\frac{\mathcal{B}-
\mathcal{B}_{0}}{\omega_{3}}\right)\right]\right)^{\psi}-
\left[\frac{n+2}{3}(\mu\psi)\right]^{\frac{\psi}{1-\psi}}\right].
\end{eqnarray}
The scalar field is
\begin{equation}\label{61}
\mathcal{B}=\mathcal{B}_{0}+\omega_{3}\Gamma\exp\left[\left(\frac{3}{n+2}\right)\frac{N}{\mu}
+\left(\frac{n+2}{3}(\mu\psi)\right)
^\frac{\psi}{1-\psi}\right]^{\frac{1}{\psi}}.
\end{equation}

The power spectrums in scalar and tensor forms can be found as
\begin{eqnarray}\nonumber
\Delta^{2}_{R}(k)&=&\left(\frac{\eta^{3}_{1}(n+2)^{\frac{9}{2}}
3^{\frac{1}{2}}(\mu\psi)^{\frac{3}{2}}}{972(1+2n)^{\frac{3}{2}}
(4\pi)^{3}(2C_{\gamma})^{\frac{1}{2}}}\right)^{\frac{1}{2}}
\left[\left(\frac{3}{n+2}\right)\frac{N}{\mu}
\right.\\\nonumber&+&\left.\left(\frac{n+2}{3}(\mu\psi)\right)
^\frac{\psi}{1-\psi}\right]^{\frac{3(\psi-1)}{4\psi}},
\\\nonumber\Delta^{2}_{T}(k)&=&\frac{2(1+2n)}{3\pi^2}(\mu\psi)^2
\left[\left(\frac{3}{n+2}\right)\frac{N}{\mu}
+\left(\frac{3}{n+2}(\mu\psi)\right)
^\frac{\psi}{1-\psi}\right]^{\frac{2(\psi-1)}{\psi}}
\\\label{63}&\times&\exp\left[-2\left(\left(\frac{3}{n+2}\right)\frac{N}{\mu}
+\left(\frac{3}{n+2}(\mu\psi)\right)
^\frac{\psi}{1-\psi}\right)^{\frac{1}{\psi}}\right].
\end{eqnarray}
The terms $n_{s}$ and $n_{T}$ related to the above equations become
\begin{eqnarray}\label{64}
n_{s}-1&=&\frac{9(1-\psi)}{4\mu\psi(n+2)}
\left[\left(\frac{3}{n+2}\right)\frac{N}{\mu}
+\left(\frac{n+2}{3}(\mu\psi)\right)
^\frac{\psi}{1-\psi}\right]^{-1},\\\nonumber
n_{T}&=&-\frac{6}{(n+2)\mu\psi}\left[\left(\frac{3}{n+2}\right)\frac{N}{\mu}
+\left(\frac{n+2}{3}(\mu\psi)\right)
^\frac{\psi}{1-\psi}\right]^{\frac{1-\psi}{\psi}}.
\end{eqnarray}
Figure \textbf{7} shows similar behavior as for the large values of
$\psi$ gives small values of the $N$. The value $N\approx20$
coincides with $n_{s}=0.96$ and verifies the compatibility of this
case with observational data. Curves for $\psi=50$ and $\psi=70$
overlap each other in the allowed range. Consequently, $R$ can be
represented as
\begin{eqnarray}\nonumber
R&=&\left(\frac{3888(1+2n)^{\frac{7}{2}}(4\pi)^3(2C_{\gamma})
^{\frac{1}{2}}(\mu\psi)^{\frac{5}{2}}}{81\pi^4\eta^{3}_{1}(n+2)^{\frac{9}{2}}
3^{\frac{5}{2}}}\right)^{\frac{1}{2}}
\left[\left(\frac{3}{n+2}\right)
\frac{3(\psi-1)}{4\mu\psi(1-n_{s})}\right]
^{\frac{5(\psi-1)}{4\psi}}\\\label{66}&\times&
\exp\left[-2\left(\left(\frac{3}{n+2}\right)
\frac{3(\psi-1)}{4\mu\psi(1-n_{s})}\right)^{\frac{1}{\psi}}\right].
\end{eqnarray}
Figure \textbf{8} verifies the compatibility of the anisotropic BI
universe model in the constant logamediate inflation regime with
WMAP7 data.
\begin{figure}
\center\epsfig{file=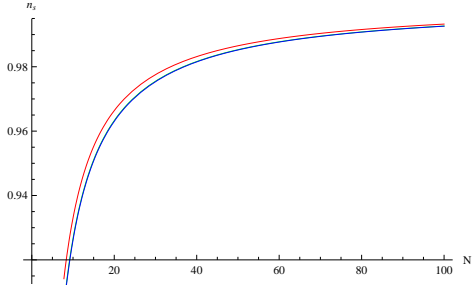, width=0.45\linewidth}\caption{The
graph of $n_{s}$ versus number of e-folds for $\mu=1,~\psi=10$(red),
$50$(green), $70$(blue), $n\approx0.5$.}
\end{figure}
\begin{figure}
\center\epsfig{file=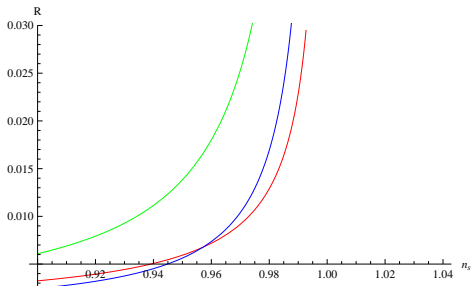, width=0.45\linewidth}\caption{The
graph of scalar-tensor ratio versus $n_{s}$ for
$\mu=1,~C_{\gamma}=70,~\psi=10$(red), $50$(green), $70$(blue),
$n\approx0.5,~\eta_{1}\propto C_{\gamma}^{\frac{1}{6}}$.}
\end{figure}

\section{Concluding Remarks}

It is well-known that inflation due to scalar fields is compatible
with completely isotropic universe. In this paper, we assume vector
field inflation which can provide the isotropic (by orthogonal
triplet vector fields) as well as anisotropic universe (with
randomly oriented $N$ vector fields). The anisotropic fields can be
achieved in two ways, i.e., by applying initial conditions on the
potential and fine-tuning of the inflationary time period. The
anisotropy is of the order $1/\sqrt{N}$ at the end of inflation. We
have investigated the role of warm vector inflation in the framework
of LRS BI universe model. A triplet of mutually orthogonal vectors
is introduced to remove the off-diagonal terms from the components
of the energy-momentum tensor as the considered model is symmetric.
We construct the field equations and conservation equations under
the slow-roll approximation using a specific form of the dissipation
coefficient. We have assumed the dissipation of the inflation
density into radiation density $(\eta>0)$.

The slow-roll parameters $(\epsilon, \lambda)$ are presented in the
context of anisotropic warm vector inflation. Using these parameters
and applying strong dissipative regime, we have calculated the more
general conditions for the starting and ending of the inflationary
era (\ref{14}). In warm inflation, thermal fluctuations are produced
instead of quantum fluctuations which are characterized by power
spectra $(\Delta^{2}_{R}(k), \Delta^{2}_{T}(k))$ and spectral
indices $(n_{S}, n_{T})$. We have evaluated all these perturbed
parameters under slow-roll conditions and finally an important
parameter, i.e., tensor-scalar ratio which is constrained by WMAP7
observations.

We have developed our model in intermediate and logamediate models
as they represent exact cosmological solutions. Each era is
discussed for two possible choices of $\eta$, i.e., it is a function
of scalar field and some positive constant. The directional Hubble
parameter, slow-roll as well as perturbation parameters are
calculated in these frameworks of inflation for anisotropic
universe. According to the observations of WMAP7, the best fit
values of these parameters $(\Delta^{2}_{R}(k),n_{S},R)$ have been
calculated with higher degree of accuracy. We have checked the
physical compatibility of our model with the results of WMAP7, i.e.,
the standard value $n_{S}=0.96$ must be found in the region
$R<0.22$. The behavior of these parameters are checked through
graphs of $N$ and $R$ versus $n_{S}$ in each case.

During intermediate scenario with variable $\eta$, we have seen that
$n_{S}=0.96$ corresponds to $N=50$. The left $R-n_{S}$ trajectories
in Figure \textbf{2} shows the inconsistency of the model with WMAP7
for $\mu=1,~C_{\gamma}=70,~g=\frac{1}{2},~\mathcal{B}_{0}\propto
C_{\gamma}^{-\frac{1}{12}},~n\approx0.5,~\eta_{0}=0.25,~1,~4$. The
method of fine-tuning helps to find the allowed range and hence
$n_{S}=0.96$ is located in the region $R<0.22$ for
$\mu=5,~g=\frac{1}{2},~\mathcal{B}_{0}\propto
C_{\gamma}^{-\frac{1}{12}},~n\approx0.5,~\eta_{0}=0.25,~1,~4$. The
isotropic universe \cite{19} for this case is compatible with WMAP7
only for $\eta_{0}=1$ keeping the same rest of the parameters. For
constant $\eta$, the model remains compatible with observational
data. Hence the anisotropic universe model is compatible with WMAP7
data during variable and constant intermediate inflation. Figures
\textbf{5}-\textbf{8} indicate the compatibility of the variable as
well as constant logamediate inflationary scenario in the framework
of anisotropic universe with WMAP7 data. We would like to mention
here that this model is consistent for all chosen values of $\psi$
and $\eta_{0}$ while isotropic universe for variable dissipation
factor is compatible only for $\eta_{0}=1$. We have also observed
that the compatibility of the model disturbs for large values of the
anisotropic parameter $(n)$. It is interesting that all the results
reduce to the isotropic universe (FRW) for $n=1$.


\begin{thebibliography}{40}

\bibitem{1} Perlmutter, S. et al.: Astron. Soc.
\textbf{29}(1997)1351; Nature \textbf{391}(1998)51; Riess, A.G. et
al.: Astron. J. \textbf{116}(1998)1009; Bennett, C.L. et al.:
Astrophys. J. Suppl. \textbf{148}(2003)1; Peebles, P.J.E. and Ratra,
B.: Rev. Mod. Phys. \textbf{75}(2003)559; Tegmark, M. et al.: Phys.
Rev. D \textbf{69}(2004)03501; Spergel, D.N. et al.: Astrophys. J.
Suppl. \textbf{170}(2007)377.

\bibitem{4} Darabi, F.: arXiv:1107.3307.

\bibitem{5} Ratra, B. and  Peebles, P.J.E.: Phys. Rev. D
\textbf{37}(1998)3406;  Chiba, T., Okabe, T. and Yamaguchi, M.:
Phys. Rev. D \textbf{62}(2000)023511; Sen, A.: J. High Energy Phys.
\textbf{48}(2002)204; Padmanabhan, T.: Phys. Rev. D
\textbf{66}(2002)021301; Guo, Z.K. et al.: Phys. Lett. B
\textbf{608}(2005)177; Padmanabhan, T.: Gen. Relativ. Gravit.
\textbf{40}(2008)529.

\bibitem{6} Gorini, V., Kamenshchik, A. and Moschella, U.: Phys. Rev. D
\textbf{67}(2003)063509; Wang, B., Gong, Y.G. and Abdalla, E.: Phys.
Lett. B \textbf{624}(2005)141; Hu, B. and Ling, Y.: Phys. Rev. D
\textbf{73}(2006)123510; Setare, M.R.: J. of Cosmology and
Astrophys. \textbf{0701}(2007)023.

\bibitem{7} Guth, A.: Phys. Rev. D \textbf{23}(1981)347; Albrecht, A. and
Steinhardt, P. J.: Phys. Rev. Lett. \textbf{48}(1982)1220.

\bibitem{8} Gold, B. et al.: arXiv:1001.4555;
Komatsu, E. et al.: Astrophys. J. Suppl. \textbf{192}(2011)18;
Larson, D. et al.: Astrophys. J. Suppl. \textbf{192}(2011)16.

\bibitem{9} Golovnev, A., Mukhanov, V. and Vanchurin, V.: J. Cosmol. Astropart.
Phys. \textbf{0806}(2008)009.

\bibitem{10} Bardeen, J.M., Steinhardt, P.J.
and Turner, M.S.: Phys. Rev. D \textbf{28}(1983)679; Linde, A.:
Phys. Lett. B \textbf{129}(1983)177; Kolb, E.W. and Turner, M.S.:
\textit{The Early Universe}, (Addison-Wesley, New York, 1990).

\bibitem{11} Bassett, B.A., Tsujikawa, S. and Wands, D.: Rev. Mod. Phys.
\textbf{78}(2006)537.

\bibitem{12} Berera, A.: Phys. Rev. Lett. \textbf{75}(1995)3218;
Phys. Rev. D \textbf{55}(1997)3346.

\bibitem{13}  Moss, I.G.: Phys. Lett. B \textbf{154}(1985)120;
Berera, A.: Nucl. Phys. B \textbf{585}(2000)666; Hall, L.M.H., Moss,
I.G. and Berera, A.: Phys. Rev. D \textbf{69}(2004)083525.

\bibitem{14} Sanyal, A.K.: Phys. Lett. B \textbf{645}(2007)1.

\bibitem{15} Yokoyama, J. and Maeda, K.: Phys. Lett. B \textbf{207}(1988)31.

\bibitem{16} Barrow, J.D.: Class. Quantum Grav. \textbf{13}(1996)2965.

\bibitem{17} Barrow, J.D.: Phys. Rev. D \textbf{51}(1995)2729 .

\bibitem{18} Ferreira, P.G. and Joyce, M.: Phys. Rev. D
\textbf{58}(1998)023503; Peebles, P.J.E. and Ratra, B.: Rev. Mod.
Phys. \textbf{75}(2003)559; Barrow, J.D. and Nunes, N.J.: Phys. Rev.
D \textbf{76}(2007)043501.

\bibitem{19} Setare, M.R. and Kamali, V.: arXiv:1309.2452.

\bibitem{20} Collins, C.B.: Phys. Lett. A \textbf{60}(1977)397;
Sharif, M. and Zubair, M.: Astrophys. Space Sci.
\textbf{330}(2010)399.

\bibitem{21} Armendariz-Picon, C.: J. Cosmol. Astropart. Phys.
\textbf{07}(2004)007.

\bibitem{22} Mukhanov, V.: \textit{Physical Foundations of Cosmology}
(Cambridge University Press, 2005).

\bibitem{23} Bastero-Gil, M., Berera, A. and Ramos, R.O.:
J. Cosmol. Astropat. Phys. \textbf{1109}(2011)033; Bastero-Gil, M.,
Berera, A., Ramos, R.O. and Rosa, J.G.: J. Cosmol. Astropart. Phys.
\textbf{1301}(2013)016.

\bibitem{24} Hwang, J.C. and Noh, H.: Phys. Rev. D
\textbf{66}(2002)084009.

\bibitem{25} Guth, A.H. and Pi, S.Y.: Phys. Rev. Lett. \textbf{49}(1982)1110;
Hawking, S.W.: Phys. Lett. B \textbf{115}(1982)295; Starobinsky,
A.A.: Phys. Lett. B \textbf{117}(1982)175; Bardeen, J.M.,
Steinhardt, P.J. and Turner, M.S.: Phys. Rev. D
\textbf{28}(1983)679.

\bibitem{26} Freese, K., Frieman, J.A. and Olinto, A.V.:
Phys. Rev. Lett. \textbf{65}(1990)3233.

\end{thebibliography}
\end{document}